\documentclass[10pt,twocolumn,letterpaper]{article}

\usepackage[spanish,english]{babel}
\usepackage[utf8x]{inputenc}
\usepackage[T1]{fontenc}
\usepackage{cite}

\usepackage[a4paper,top=3cm,bottom=2cm,left=3cm,right=3cm,marginparwidth=1.75cm]{geometry}

\usepackage{amsmath}
\usepackage{graphicx}
\usepackage{booktabs}
\usepackage[colorinlistoftodos]{todonotes}
\usepackage[colorlinks=true, allcolors=black]{hyperref}

\title{
		\usefont{OT1}{bch}{b}{n}
		\normalfont \normalsize \textsc{} \\ [10pt]
		\huge A Minimal-Input Multilayer Perceptron for Predicting Drug-Drug Interactions Without
		Knowledge of Drug Structure\\
}

\usepackage{authblk}
\author[0]{Alun Stokes, William Hum, Jonathan Zaslavsky \\
		McMaster University \\}

\begin{document}
\maketitle

\selectlanguage{english}
\begin{abstract}
The necessity of predictive models in the drug discovery industry cannot be understated. With the sheer volume of potentially useful compounds that are considered for use, it is becoming increasingly computationally difficult to investigate the overlapping interactions between drugs. Understanding this is also important to the layperson who needs to know what they can and cannot mix, especially for those who use recreational drugs - which do not have the same rigorous warnings as prescription drugs. Without access to deterministic, experimental results for every drug combination, other methods are necessary to bridge this knowledge gap. Ideally, such a method would require minimal inputs, have high accuracy, and be computationally feasible. We have not come across a model that meets all these criteria. To this end, we propose a minimal-input multi-layer perceptron that predicts the interactions between two drugs. This model has a great advantage of requiring no structural knowledge of the molecules in question, and instead only uses experimentally accessible chemical and physical properties - $20$ per compound in total. Using a set of known drug-drug interactions, and associated properties of the drugs involved, we trained our model on a dataset of about $650,000$ entries. We report an accuracy of $0.968$ on unseen samples of interactions between drugs on which the model was trained, and an accuracy of $0.942$ on unseen samples of interactions between unseen drugs. We believe this to be a promising and highly extensible model that has potential for high generalized predictive accuracy with further tuning.
\end{abstract} 

\section{Introduction}

With the rapid technological advances in the drug discovery process as well as the prevalence of increasingly-many drugs in our lives, understanding the ways in which drugs affect one another in the body is a pressing problem. Not only do we need to delineate individual effects of chemicals in the body, we must be able to identify the potential for interaction in order to better design these compounds. Exhaustively testing every drug-drug combination is both cost and time-prohibitive due to the sheer volume of compounds to be tested, making optimized methods necessary to perform proper analysis \cite{kastrin_predicting_2018}. A practical approach for identifying potential drug interactions and assessing associated risks has emerged in recent years with the development of predictive pharmacological networks \cite{cami_pharmacointeraction_2013}. The applications of such methods are far reaching in both the process of drug discovery, as well as the societal good that would come with easy, dynamic access to tools that allow laypeople to better control their health. The prevalence of the use of recreational drugs sees many drug-related ailments coming as a result of mixing drugs that should not be mixed. As such, developing accessible, fast models to determine these interactions is of great importance to our society \cite{kastrin_predicting_2018}. In this paper, we propose the use of machine learning methods and known drug-drug interactions to predict these interactions in unknown drug pairings, as well as drugs on which a model has not been trained.

One of the largest barriers to predicting drug-drug interactions is the complexity of factors associated with that interaction. As such, many deterministic or exhaustive methods fail the test of feasibility due to their necessary computational complexity. Even such a problem as codifying the molecular structure of a chemical leads to a great complexity of information to be processed, with this processing being non-trivial of its own merit. Therefore, any strategy to predict these interactions reasonably and at scale needs to be able to do so on a minimal set of parameters, with limited computational complexity. Our model fits these requirements given each drug molecule is parameterized by 20 floating-point values, and all the necessary computation consists of matrix operations that are hardware-accelerated on modern computers. This leads to a model which, on an average consumer laptop processor (Intel\textsuperscript{\textregistered} Core i5-8259U), can predict interactions for over 10,000 drug pairs per second - with $\approx 0.968$ accuracy.

This model determines the binary condition of interactivity, rather than categorizing the type or location of interaction between two drugs. The implementation of the latter is, however, expected to progress naturally from the presented model, and would only require superficial adjustments. The reason for using a less specific model in this paper is two-fold: the limited accessibility of a dataset that characterizes the location and types of interactions, as well as the amount of time allotted for completion of this project. We, however, do not feel that this is a weakness or limitation of the model.

The drug-drug interaction dataset used in this paper was provided by The Science and Engineering of Emerging Systems Lab \cite{guimera_network_2013}, who collected their data from DrugBank. The molecular properties of each drug were obtained from PubChem \cite{kim_pubchem_2019}, a database maintained by the National Centre for Biotechnology Information. 

\section{Materials \& Methods}

The primary dataset used in this paper \cite{guimera_network_2013}, is a line-separated compilation of lists of drugs that are known to interact with one another. There are approximately $1,200$ unique species documented, and about $225,000$ individual drug-drug interactions. Certain parts of the dataset were unusable, as they specified macromolecule-drug interactions, and we are only interested in small-molecule drugs. Additionally, certain compounds did not have entries in \cite{kim_pubchem_2019} that could be programmatically downloaded and processed, so those interactions were discarded. This left about $190,000$ drug-drug interactions for the training, validation, and testing of the model. A larger and more generalized set would do well to help the model train to determine the important factors in a drug-drug interaction. As such, several data augmentation techniques were performed on the set in order to increase its size, whilst not perverting the accuracy or validity of its information.

First, given we only had records of drug-drug interactions, we needed to generate a set of drug-drug non-interactions. Given that this dataset lists every subset of the drugs that do interact, a list of non-interacting drugs could be generated by permuting two drugs in the set, and checking for their inclusion in the original dataset. This doubled the size of the original dataset - although it should be noted that doubling was chosen to keep the training categories evenly distributed. This improves training performance and avoids the necessity for weighting the training rates per category. Next, each 2-tuple of interacting drugs was reversed and added back into the list, again doubling the dataset. The order in which the two drugs are fed into the model should be independent of their interaction, and as such it should be trained to find an interaction between two drugs, regardless of which is first in the list. This brought the size of the dataset up to approximately $760,000$ drug pairs. This set was then shuffled in order to prevent the model from learning any patterns present in the original presentation of the data. Of this full set, $10\%$ is set aside for validation during training, and a further $10\%$ composes the testing set. These are chosen randomly, although for the purposes of comparing models, the generator seed was fixed in order to allow comparable statistics.

Once the interaction set was processed, information of the individual compounds needed to be associated with each drug in an interaction. This data was obtained from \cite{kim_pubchem_2019}, and certain attributes were extracted and assigned to the interacting molecules to be fed into the model. The full list of extracted properties can be see in Table \ref{inputs-table}. Note that although these are mostly integer-valued parameters, for the purposes of computation they are treated as floating-point values.

\begin{table}
    \caption {Chemical Parameters for Input into Model}
    \label{inputs-table}
    \begin{center}
        \begin{tabular}{@{}lllll@{} p{3cm}}
        \toprule
        Parameter                                         & Type    &  &  &  \\ \midrule
        \# of C atoms                                     & Integer &  &  &  \\
        \# of H atoms                                     & Integer &  &  &  \\
        \# of N atoms                                     & Integer &  &  &  \\
        \# of O atoms                                     & Integer &  &  &  \\
        \# of F atoms                                     & Integer &  &  &  \\
        \# of S atoms                                     & Integer &  &  &  \\
        \# of P atoms                                     & Integer &  &  &  \\
        \# of Cl atoms                                    & Integer &  &  &  \\
        \# of I atoms                                     & Integer &  &  &  \\
        \# of Br atoms                                    & Integer &  &  &  \\
        Log (P)                                           & Decimal &  &  &  \\
        \# of H-bond acceptors                            & Integer &  &  &  \\
        \# of H-bond donors                               & Integer &  &  &  \\
        \# of rotatable bonds                             & Integer &  &  &  \\
        \# of heavy atoms                                 & Integer &  &  &  \\
        Complexity (Bertz formula)                        & Integer &  &  &  \\
        \# of def. atom stereocentres                     & Integer &  &  &  \\
        \# of undef. atom stereocentres                   & Integer &  &  &  \\
        \# of def. bond stereocentres                     & Integer &  &  &  \\
        \# of undef. bond stereocentres                   & Integer &  &  &  \\ \bottomrule
        \end{tabular}
    \end{center}
\end{table}

The neural network is a feed-forward network, meaning propagation only goes in the forward direction. Specifically, it is a multi-layer perceptron, with numerous hidden layers. Given the assumed non-linearity of this problem, at least several latent layers were necessary to predict interactions with any degree of accuracy. Each latent layer is either simply a fully-connected (FC) layer, a dropout layer, or a batch normalization (BN) layer. The network follows a diamond-like structure for the number of nodes per layer, starting with $256$, increasing over several layers to $512$, and then decreasing over several layers to $128$ before the final output. Every FC layer is proceeded by a BN layer. Table \ref{layers-table} breaks down each layer of the network. Additionally, layers are grouped together to create a `super-layer' structure composed of two identically sized sets of an FC and BN layer, all followed by a dropout layer. The purpose of these dropout layers is to prevent the model from being able to `memorize' the training data to artificially increase its training accuracy. Both the first and second latent layers as well as the last latent layer lack a dropout layer proceeding them to prevent early-stage data loss and highly variable classification respectively. The FC layers use rectified linear units (ReLUs) for activation, while the output layer uses a sigmoid function. Adam was used as the optimizer, with parameters left as default. All dropout layers use a drop rate of $0.3$, and the total number of trainable parameters in the model is $744,449$. Various numbers of epochs were used for training in order to identify the optimal value, and batch sizes between 32 and 256 were tested for the same purpose.

After training, the model is evaluated on the testing set to quantify how well it has learned - i.e. its generalizability.  As well, the model is tested on the interactions (or lack thereof) of drugs upon which it has not been trained - a further test of how well this model actually predicts drug-drug interactions. The data for this last test is obtained from \cite{kim_pubchem_2019}.

\begin{table}
    \caption {Layers Composing Model}
    \label{layers-table}
    \begin{center}
        \begin{tabular}{@{}llll@{}}
            \toprule
            Type    & Activation & Output & Parameters \\ \midrule
            Input   &            & 40     &            \\
            FC      & ReLU       & 256    & 11520      \\
            BN      &            & 256    & 1024       \\
            FC      & ReLU       & 256    & 65792      \\
            BN      &            & 256    & 1024       \\ \midrule
            FC      & ReLU       & 512    & 131584     \\
            BN      &            & 512    & 2048       \\
            FC      & ReLU       & 512    & 262656     \\
            BN      &            & 512    & 2048       \\
            Dropout &            & 512    &            \\ \midrule
            FC      & ReLU       & 256    & 131328     \\
            BN      &            & 256    & 1024       \\
            FC      & ReLU       & 256    & 65792      \\
            BN      &            & 256    & 1024       \\
            Dropout &            & 256    &            \\ \midrule
            FC      & ReLU       & 128    & 32896      \\
            BN      &            & 128    & 512        \\
            FC      & ReLU       & 128    & 16512      \\
            BN      &            & 128    & 512        \\
            Dropout &            & 128    &            \\ \midrule
            FC      & ReLU       & 128    & 16512      \\
            BN      &            & 128    & 512        \\
            FC      & Sigmiod    & 1      & 129        \\ \bottomrule
        \end{tabular}
    \end{center}
\end{table}

\section{Results}
In this project, the architecture of the network remained fundamentally unchanged throughout testing after being determined from early tests on several variants. These early tests were focused on the number of FC layers in each super-layer, as well as the number of super-layers (i.e. the depth of the network). As well, it was determined how many FC layers were to be included before the first super-layer, and after the final. As can be seen in Table \ref{layers-table}, it was found that the model performed optimally (in terms of the validation accuracy after 100 epochs) with two FC layers, then three super-layers, then a single FC layer before the output. In whole, every combination of 1 to 3 FC layers before the super-layers, 1 to 5 super-layers, and 1 to 3 FC layers after the super-layers was tested. As well, different numbers of neurons in each layer and super-layer were tested, all with the diamond pattern described earlier. It was found that the sizes 256, 512, 256, 128, 128 performed best of the tested hyperparameters.

\begin{figure}[htb]
    \centering
    \includegraphics[width=8cm]{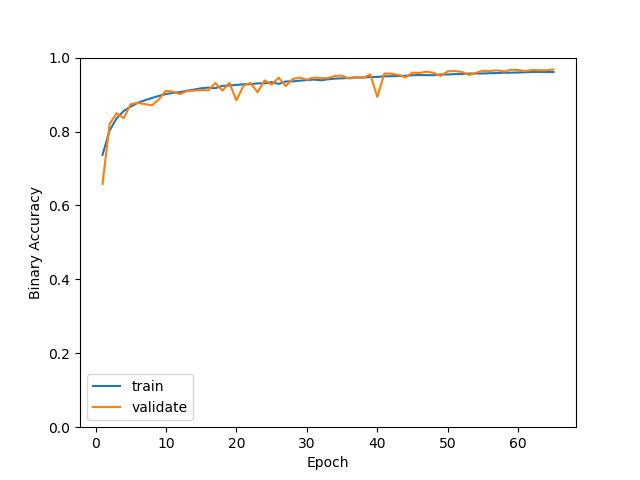}
    \caption{A plot of the training and validation accuracy over $65$ epochs for a model trained on approximately $650,000$ samples with a batch size of $64$. Training took about $120$ minutes. The testing accuracy for this model is 0.965 on unseen interactions, and 0.942 on unseen drugs.}
    \label{figure-val-acc}
\end{figure}

\begin{figure}[htb]
    \centering
    \includegraphics[width=8cm]{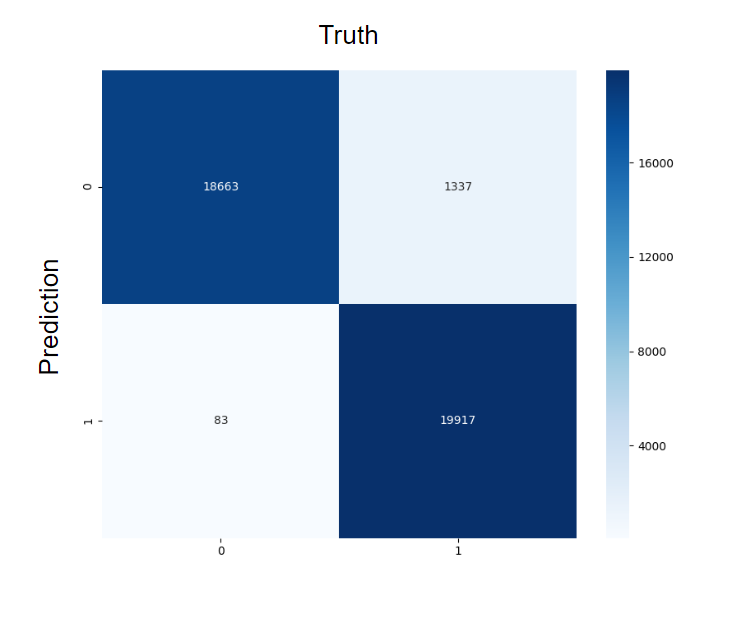}
    \caption{A confusion matrix showing the correct and incorrect predictions of drug-drug interactions over the seen drugs testing set. The column-wise contrast of colour intensity represent the statistical power and specificity of $0.937$ and $0.996$ respectively.}
    \label{figure-conf-mat}
\end{figure}

The final, best performing model used the architecture mentioned previously, with uniform dropout rates of $0.3$, a batch size of $64$, and trained over $65$ epochs. The training accuracy curve can be seen in Figure \ref{figure-val-acc}. Of particular note is how well the validation accuracy mimics the training accuracy. When tested on a set of unseen interactions for known drugs, the model scored a binary accuracy of $0.968$, lining up very closely with the validation accuracy at epoch $65$ of $0.965$. When tested on a set of unseen interactions between unknown drugs, the model scored a binary accuracy of $0.942$. This small drop in accuracy is expected due to the generalization of the data being tested on. A confusion matrix for the can be seen in Figure \ref{figure-conf-mat}. The model itself is relatively small, with the whole architecture and learned weights coming out to about $9$MB in size.

\section{Discussion}
In training a slew of models based on the determined architecture, we were able to tune the hyperparameters - including the dropout rate, output size of FC layers, activation functions and learning rate - and ascertain the best performance from our chosen architecture. Although simplistic, the simplicity is what makes the model as efficient as it is. To train over $65$ epochs on about $650,000$ interactions takes just under two hours on two Nvidia\textsuperscript{\textregistered} P100 GPUs - quite a short time in this field. This makes the model highly extensible to much larger datasets, while staying within reasonable training times, even without optimal hardware. The potential for models such as this to find a purpose outside dedicated research laboratories hinges on their efficiency. To be used in a consumer-facing application that predicts drug interactions, this model needs to be accessible outside of the field - in terms of necessary computational power, ease of use, and ease of access to necessary information. We believe that our model meets these criteria.

Of most important note from our results is the final accuracies this model was able to achieve for both known and unknown drugs. The latter of the two was lower, at only $0.942$, but this is to be expected with the generalization of the model. The fact that the drop in accuracy was so small (just over $0.02$) is a great encouragement to the efficacy of this model at predicting drug-drug interactions. It seems to have been able to extract some of the most important information that determines whether two drugs will interact, as it does so without having seen those drugs in its training.

The importance of models as the one we have created here cannot be overstated. Being able to predict drug-drug interactions is vitally important in both the commercial setting of drug development, as well as in the context of an individual being able to know and understand what drugs they should and should not be able to take together. In order for a model to address the needs in whatever sector, it must be able to, with high accuracy, quickly determine from minimal information whether two compounds will interact. Where many other models working to this end have failed previously is in this last criterion \cite{cami_pharmacointeraction_2013, kastrin_predicting_2018}. Most models require an intimate knowledge of the structures of the chemicals of interest at a minimum, and often require in addition to that a host of other information that is not easily accessible for any given molecule. The strength of our minimal information requirement is that these parameters can be experimentally determined without knowing or having to encode the structure of the molecule, and uses very readily available properties of compounds. This saves not only on the burden of procuring datasets that have all the necessary information, it reduces the number of necessary tests to be run on a compound before one has the requisite amount of necessary information to predict the interaction of two drugs. This not only improves the logistics of using such a model, it decreases the amount of time to both train and utilize that model and makes it more accessible in whole.

As mentioned previously, one of the primary extensions of the model in its current state is to switch from the binary classification of interaction to a categorical classification of the type and location of interaction. We consider this a natural extension, and preliminary tests with small sets indicates promising results, similar to the accuracy for binary classification. The primary barrier to this is the procurement of a dataset that has labelled interaction information, but this information is available to be created through processing raw data provided by \cite{kim_pubchem_2019}.

A further improvement is to increase the number of parameters read into the network for each drug. While the set given provides a relatively high degree of accuracy, the classification accuracy could be improved by including other, easily obtainable properties of the chemicals of interest. Again, the only boundary to this is the time it would take to compile a labelled dataset with the requisite information. 

Finally, we would like to perform sensitivity analysis on the model. This is an endeavour in and of itself, but doing so would allow for an entirely new and informative purpose to the model by correlating each input with its effect on the model's output. This is effectively a way to determine which chemical properties are most important in determining whether or not they will interact in the body. The hope in doing so is to further investigate where areas of research should focus to best characterize and predict these interactions. 

\section*{Conclusions}
With available resources and sufficient drug data, this model has potential to predict drug-drug interactions with a high degree of accuracy. The implications of this model are capable of great societal benefit, in terms of pharmacoeconomics and public health \cite{kastrin_predicting_2018,takeda_predicting_2017}. With sufficient accuracy, the model can reduce high development expenses and save time by avoiding rigorous testing for drug-drug interactions during drug discovery. By predicting interactions computationally rather than experimentally, the efficiency of the pharmaceutical industry can be improved \cite{takeda_predicting_2017}. As the number of drugs and chemicals continues to grow, along with their interactions, models such as this become increasingly important.

With recreational drug use becoming more common, the importance of drug-drug interaction prediction is exceedingly pertinent \cite{kastrin_predicting_2018}. The growing variety of both legal and illegal recreational drugs raises the potential for unforeseen consequences related to various drug combinations \cite{kastrin_predicting_2018}. Using the minimal-input multi-layer perceptron network described here allows for quick prediction with sufficient information (which is largely readily available online) to predict possible problematic interactions. The availability of the model to the world of medicine as well as the public can be far reaching in improving public health.

\bibliography{main}

\end{document}